%% file: idealpolar.tex
\documentclass[11pt]{amsart}

\usepackage{calc}
\usepackage[pdftex]{color}
\usepackage[pdftex]{graphicx}
\usepackage{epstopdf}
\usepackage{epsfig}
\usepackage{amsmath}
\usepackage{amsxtra}
\usepackage{amssymb}
\usepackage{float}

\newcommand{\strictly}{}
\newcommand{\Grun}{\mathcal{G}}
\newcommand{\dehug}{\delta}
\newcommand{\creq}{\overset{\dehug\turn=0}{=}}

\newcommand{\jmp}[1]{\Delta#1}
\newcommand{\jmpl}[1]{\Delta(#1)}
\renewcommand{\vec}[1]{\mathbf{#1}}

\newcommand{\bet}{\beta}
\newcommand{\turn}{\theta}
\newcommand{\avg}[1]{\bar{#1}}

\newcommand{\Mach}{M}
\newcommand{\Machn}{M^n}

\newcommand{\uvar}{u}
\newcommand{\ua}{\uvar}
\newcommand{\uu}{\vec\uvar}
\newcommand{\ux}{\uvar^x}
\newcommand{\uy}{\uvar^y}
\newcommand{\ut}{\uvar^t}
\newcommand{\un}{\uvar^n}
\newcommand{\ja}{j}
\newcommand{\uau}{\ua_0}
\newcommand{\jau}{\ja_0}

\newcommand{\jj}{\vec j}

\newcommand{\jn}{j^n}

\newcommand{\fod}{f}
\newcommand{\Zcomp}{Z}
\newcommand{\hug}{K}
\newcommand{\Hpm}{H}
\newcommand{\hpm}{h}
\newcommand{\epm}{e}
\newcommand{\spm}{s}

\newcommand{\pp}{p}
\newcommand{\idens}{V}
\newcommand{\dens}{\varrho}

\newcommand{\temp}{T}
\newcommand{\stemp}{\temp} 
\newcommand{\csnd}{c}

\newcommand{\Rspec}{R_s}
\newcommand{\Runiv}{R_u}
\newcommand{\kB}{k_B}

\newcommand{\cpspec}{c_p}

\newcommand{\cvspec}{c_v}
\newcommand{\gisen}{\gamma}

\newcommand{\nn}{\vec n}
\newcommand{\ts}{\vec t}

\def\XXint#1#2#3{{\setbox0=\hbox{$#1{#2#3}{\int}$}
\vcenter{\hbox{$#2#3$}}\kern-.5\wd0}}

\if11%
\newcommand{\dati}[3]{(\partial#1/\partial#2)_{#3}}
\newcommand{\dat}[3]{\mathchoice{\big(\frac{\partial#1}{\partial#2}\big)_{#3}}{\dati#1#2#3}{\dati#1#2#3}{\dati#1#2#3}}
\newcommand{\pddat}[3]{\mathchoice{\big(\frac{\partial^2#1}{\partial#2^2}\big)_{#3}}{(\partial^2#1/\partial#2^2)_{#3}}{(\partial^2#1/\partial#2^2)_{#3}}{(\partial^2#1/\partial#2^2)_{#3}}}

\fi%

\newcommand{\csep}{\quad,\quad}

\newcommand{\defm}[1]{\emph{#1}}
\newcommand{\subeq}[2]{\mathord{\underbrace{\mathop{#1}}_{#2}}}
\newcommand{\supeq}[2]{\mathord{\overbrace{\mathop{#1}}^{#2}}}

\newcommand{\impl}{\Rightarrow}

\newcommand{\sign}{\operatorname{sign}}

\newcommand{\half}{\frac12}

\newcommand{\myeqref}[1]{\eqref{#1}}

\newcommand{\topref}[2]{\overset{\text{\eqref{#1}}}{#2}}

\newcommand{\dotp}{\boldsymbol{\cdot}}

\newcommand{\conv}{\rightarrow}

\newcommand{\qiq}{\quad\impl\quad}

\newcounter{eqno}[section]
\newcommand{\myeqlab}[1]{\refstepcounter{eqno}\tag{\arabic{section}.\arabic{eqno}}\label{#1}}

\begin{document}

\title{Shock polars for ideal non-polytropic gas}

\newif\ifGPblacktext
\GPblacktextfalse

\author{Volker W. Elling}
\email{velling@math.sinica.edu.tw}
\address{Institute of Mathematics, Academia Sinica, Taipei}

\begin{abstract}
  We show that shock polars for ideal non-polytropic gas (thermally but not calorically perfect)
  have a unique velocity angle maximum, the critical shock,
  assuming convex equation of state (positive fundamental derivative) and other standard conditions.
  We also show that the critical shock is always transonic. 
  In the process we show that temperature, pressure, energy, enthalpy, normal mass flux and entropy are 
  \strictly increasing along the forward Hugoniot curves,
  and hence along the polar from vanishing to normal shock; speed is \strictly decreasing along the entire polar,
  mass flux and importantly Mach number are decreasing on \emph{subsonic} parts of the polar. 
  
  If the equation of state is ideal but not convex, or convex but not ideal, counterexamples can be given with multiple critical shocks,
  permitting more than two shocks attaining the same velocity angle, in particular more than one shock of weak type. 
\end{abstract}

\maketitle
\thispagestyle{empty}

\section{Background}

\begin{figure}
    \centerline{\fbox{\footnotesize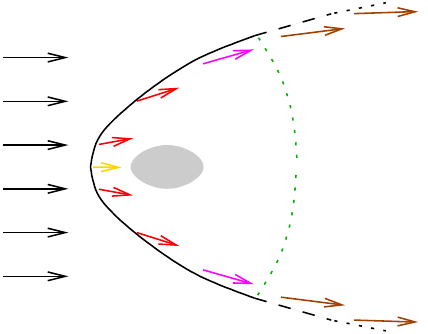}}
    \caption{Bow shock ahead of a blunt body 
      in supersonic flow 
      (subsonic regions need not enclose the body)
    }
    \label{fig:bowshock}
    \vspace{1cm}
    \centerline{\footnotesize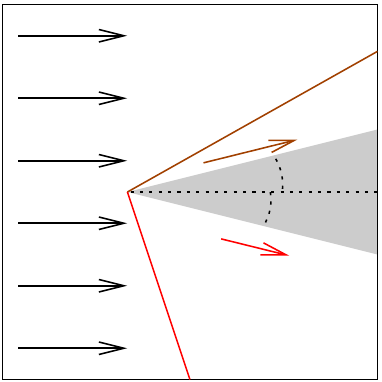}
    \caption{Supersonic flow onto wedge (strong shock usually not observed)}
    \label{fig:wedge}
    \vspace{1cm}
    \centerline{\footnotesize\input{polar.xxx}}
    \caption{Shock polar ($\fod=5$ polytropic, $\Mach_0=2.5$), symmetric across $\uu_0=(1,0)$ axis}
    \label{fig:polypolar}
\end{figure}

Given a fixed state on the upstream side of an oblique steady shock, the \defm{shock polar} (fig.\ \ref{fig:polypolar}) is the curve of all possible downstream velocities that can be generated by varying the angle of the shock. In bow shocks ahead of a blunt body
(fig.\ \ref{fig:bowshock}) all velocities along the polar are realized, with a normal shock on the symmetry axis
and the limit of vanishing shocks at infinity. 
Flow onto sharp narrow wedges (or blunt ones seen at a distance) has shocks attached to the leading edge (fig.\ \ref{fig:wedge}). 
The shock must turn the upstream velocity $\uu_0$ by a given angle $\turn$ to produce a downstream velocity $\uu$ parallel to the solid surface. 
For $\turn$ less than the \defm{critical angle} the polar (fig.\ \ref{fig:polypolar}) shows two shock solutions\footnote{%
  for larger angles, i.e.\ blunt wedges, there are no attached-shock solutions;
  the third solution labelled ``expansive'' is an unphysical expansion shock}; the weaker one is commonly observed. 
Shock polars are also important for many other flow patterns with oblique shocks, for example interaction with expansion fans, slip lines and density discontinuities, incident shocks interacting with each other or with solid surfaces, in regular or Mach reflections\footnote{\cite{neumann-1943,ben-dor-book,elling-liu-pmeyer,elling-nonexstrong,elling-rrefl-lax,skews-ashworth,vasiliev-kraiko,hunter-tesdall}}, etc.

The shock polar depends not only on the upstream state, in particular its Mach number $\Mach_0$, but also on the fluid.
We call gas \defm{ideal}\footnote{%
  sometimes called ``perfect'' or ``thermally perfect''}
if pressure $\pp$, volume per mass $\idens=1/\dens$ and temperature $\temp$ are related by $\pp\idens/\temp=\Rspec$ with constant $\Rspec$. 
Then internal energy per mass $\epm$ is a function 
of temperature alone. We call ideal gas \defm{polytropic}\footnote{sometimes called ``calorically perfect''}
if the function is linear: $\epm=\half\fod\Rspec\temp$ with constant $\fod$. 

For polytropic gas exact polar formulas have been known for a long time\footnote{\cite[part B]{meyer-fan-ramp}, \cite[sec.\ 27]{busemann-handbuch-experimentalphysik}, \cite[(121.03)]{courant-friedrichs}}: for $\uu_0$ rotated and scaled to $(\ux_0,\uy_0)=(1,0)$, 
\newcommand{\uxn}{\ux_{\text{nor}}}
\newcommand{\uxm}{\ux_{\max}}
\begin{alignat*}{5} \uy = \pm (1-\ux) \sqrt{\frac{\ux-\uxn}{\uxm-\ux}} \csep \uxn = \frac{1+\fod\Mach_0^{-2}}{\fod+1} \csep \uxm = 1 + \frac{\fod\Mach_0^{-2}}{\fod+1},  \end{alignat*}
where $\uxn$ is $\ux$ for a normal shock. Using this formula it is easily checked that the ``$+$'' branch of $\uy$ is a strictly concave function of $\ux$ between $\uxn$ and $\ux_0$, so that the upper half of the polar has exactly one local maximum of velocity angle, a \defm{critical-type} 
shock where the polar is tangent to the $\turn$ ray. No solutions exist for larger $\turn$; 
for smaller there is a \defm{weak-type} shock where the polar crosses the ray nontangentially from below to above (as we pass from vanishing to normal shock, see fig.\ \ref{fig:polypolar}), and an opposite crossing at the \defm{strong-type} shock.

Although strong-type reflections can be generated in some settings by careful adjustment of parameters\footnote{in particular $\Mach_0$, solid shapes, downstream conditions}, they tend to disappear when these parameters are perturbed, while the corresponding weak-type reflections are robust, even when transonic.\footnote{%
  Earlier suggestions that strong-type shocks are unstable because they are transonic turned out to be incorrect. 
  Already \cite{teshukov} proposed based on linearized analysis that transonic shocks are in some sense stable \emph{dynamically} 
  (under initial data perturbation) if they are weak-type.
  Shock-capturing numerics found both types dynamically stable in a natural sense (\cite[fig.\ 4]{elling-liu-rims05}), 
  but \emph{structural} instability turned out to be the correct notion.}
\cite{elling-sonic-potf,elling-detachment} give a rigorous proof in the case of certain regular reflections in nonlinear compressible full potential flow. 
Mathematically, if a problem is posed so that its linearization around a weak-type reflection has exactly one solution with bounded velocity,
then strong-type reflections generally produce unbounded velocity in the shock-solid corner;
velocities above the limit speed\footnote{which is bounded by upstream flow since total enthalpy is continuous across the shock}
are physically meaningless and cannot be used to construct small perturbations to solutions of the full nonlinear problem. 

Altogether, in any given reflection problem a weak-type shock is likely to appear,
and for polytropic polars there is only one such shock.

\section{Motivation}

However, ``polytropic'' is a rather loose approximation for the most important gases. 
Set $\fod=2(\cpspec/\Rspec-1)$ with specific heat at constant pressure $\cpspec=\dat\hpm\temp\pp$, 
where subscript $\pp$ indicates the $\pp$ coordinate is held constant for taking the $\temp$ partial derivative;
$\hpm=\epm+\pp\idens$ is enthalpy per mass. 
Polytropic flow is characterized by constant $\fod$; $\fod=3$ is used for monatomic gases, $\fod=5$ for diatomic ones like oxygen. 

For oxygen at atmospheric pressure, $\fod$ stays between 5.0 and 5.1 up to room temperature (see fig.\ \ref{fig:zfo}), but crosses above 5.4 already at 500 Kelvin. 
Similar observations can be made for most other multiatomic gases, including the important cases of nitrogen/air, carbon dioxide, and hydrogen\footnote{cf.\ fig.\ \ref{fig:foh}, see also \cite[fig.\ 3]{woolley-scott-brickwedde-1948} for experimental/modelled $\cpspec$ at high temperature}. 

Clearly $\fod$ is near-constant only over narrow temperature ranges, which are easily exceeded by many applications,
e.g.\ high-supersonic flight, flow after combustion, ramjet/scramjet inlets, 
or atmospheric entry (\cite{boyd-candler-levin-pof1995,gnoffo-1999}). 

In classical mechanics $\fod$ represents degrees of freedom per gas particle, three for linear motion of hard-sphere models of atoms, 
another two for visible rotation axes of rigid-dumbbell models of diatomic molecules etc., 
with equal average energy $\half\kB\temp$ in each degree (Boltzmann equipartition),
and a continuum of possible energies. 
But energy levels are discrete in quantum-mechanical reality. There, constant integer $\fod$ are only suitable when 
the quanta are either much larger than temperature, so that most molecules are in the ground state (degree of freedom effectively absent, as for bond-axis vibration of oxygen or nitrogen at room temperature), or much smaller so that the energy levels resemble a continuum (rotational quanta for most gases, with hydrogen a notable exception (cf.\ fig.\ \ref{fig:foh})).
At intermediate temperatures $\fod$ necessarily rises through fractional values. 

\begin{figure}
    \centerline{\scriptsize\input{o2woo.xxx}}
    \caption{$\Zcomp$ (lower curves) and $2(\cpspec/\Rspec-1)$ (upper) for molecular oxygen (\cite{woolley-1953}). $\Zcomp\approx 1$ indicates ideal behaviour, constant $2(\cpspec/\Rspec-1)$ polytropic. }
    \label{fig:zfo}
    \centerline{\scriptsize\input{h2wsb.xxx}}
    \caption{$2(\cpspec/\Rspec-1)$ for molecular hydrogen (normal); rotational quanta are large enough for decrease to $3$ at low $\temp$ (\cite{woolley-scott-brickwedde-1948})}
    \label{fig:foh}
\end{figure}

\begin{figure}
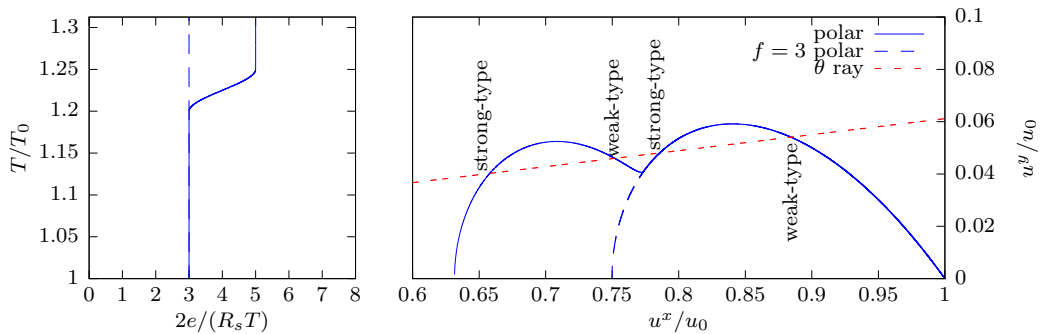

  \centerline{\scriptsize\input{plotidealf.xxx}%
  \input{plotidealpolar.xxx}}%
  \caption{$\Mach_0=1.225$ ideal gas polar with multiple velocity angle maxima, 
    caused by a cubic spline increase in $\epm/\temp$.}
  \label{fig:idealdent}
\end{figure}

Beyond polytropic, the next more accurate model is non-polytropic ideal gas.
But there we can find examples of $\epm=\epm(\temp)$ that have shock polars with multiple velocity angle maxima
(see fig.\ \ref{fig:idealdent}, section \ref{section:idealpolar}),
so that some turning angles $\turn$ have four or more theoretical solutions.
In order of increasing strength\footnote{%
  the standard terms ``weak'' and ``strong'' are especially prone to confusion here}
the solutions alternate between strong and weak type. 
Even if we exclude the strong-type half on stability grounds, multiple weak-type shocks remains. 

Multiple solutions are a major concern for applications.
  Regular and Mach reflections are already suspected to coexist for some parameters, with steady flow switching between them in unsteady hysteresis
  (\cite{hornung-oertel-sandeman,chpoun-passerel-li-bendor-1995experimental,laguarda-hickel-schrijer-oudsheusden-2020}). 
  Multiple solutions can cause numerics or experiments to overlook flows that occur in reality. 
  Such mispredictions can have significant effects, persistent like incorrect stress or heat flow estimates, or spontaneous like engine unstart, possibly hard to diagnose even in hindsight.
  Even when all steady flows are detected in advance, spontaneous and unpredictable transitions between them are unacceptable in many applications
  (consider sudden flow changes or vibrations in engines operating close to their stress/temperature limits).

One route is to look for reasons why one of several weak-type shocks should be preferred.
  However, considering that the change of $\epm(\temp)$ in fig.\ \ref{fig:idealdent} is unusually sharp,
  we should first look for simple conditions excluding the possibility of multiple weak-type solutions altogether.

They can easily be excluded for any concrete $\epm(\temp)$ and $\temp_0,\pp_0,\Mach_0$ by numerical calculations. 
Numerics are commonly used to compute polars or Hugoniot curves in complex non-ideal cases (e.g.
  \cite[fig.\ 6 and 7]{short-quirk-jfm2018}, \cite[fig.\ 1]{alferez-touber-jfm2017}, \cite[section 6.1]{lieberthal-stewart-hernandez-jfm2017}, \cite[fig.\ 1]{huete-vera-2019}, \cite[fig.\ 4]{vimercati-gori-spinelli-guardone});
  in such complex cases there is little hope to find simple analytic formulas for the shock polar.
  However, numerical plotting has to be repeated for each fluid and each upstream state.
  Even for ideal gas the possible equations of state form an infinite-dimensional parameter space; trying to cover it numerically is slow and error-prone.

Nor is it necessary: for ideal gas we show 
that the allowed $\epm(\temp)$ can be characterized precisely by the familiar condition of convex equation of state\footnote{also expressed as ``positive fundamental derivative''}, namely
positive $\pddat\pp\idens\spm$. 
This is satisfactory since only a few materials are known or claimed to violate the condition, 
in regions far from ideal behaviour.\footnote{%
  Besides, whenever it is violated compressive shocks may be non-admissible, so that the correct flow turning 
  by a particular angle $\turn$ may well be a composite of several waves rather than a single shock.}
Our other conditions are standard: positive heat capacity, positive speed of sound etc.\ (see section \ref{section:assumptions}).

Idealness is a close approximation over large temperature ranges. 
This is quantified using the \defm{compressibility factor} 
\begin{alignat*}{5} \Zcomp = \frac{\pp\idens}{\Rspec\temp}, \end{alignat*}
with $\Zcomp$ near-constant\footnote{As in \cite{woolley-1953} we use $\Rspec=\Runiv/M$ where $M$ is molar mass at a reference state such as standard conditions; 
  as temperature increases $\Zcomp$ rises through dissociation of molecular to atomic oxygen from $1$ to a new plateau of $2$ which is again considered ideal.} considered ideal behaviour.
For oxygen at atmospheric pressure $\Zcomp\approx 1$ from $100^\circ$K to well over $3000^\circ$K (fig.\ \ref{fig:zfo}). 
For most gases the low end of the ideal temperature range is due to 
increasing density permitting significant intermolecular forces, 
usually near the boiling/sublimation point ($\approx$ $90^\circ$K for oxygen at 1 atm) except at high pressures. 
The high end is usually due to dissociation or ionization, e.g.\ $\text{O}_\text{2}\rightleftharpoons 2\text{O}$ between $3000$--$4000^\circ$K, 
higher for nitrogen. However, a fully dissociated neutral gas is ideal again.\footnote{$\pp=\Runiv\temp n$ ($n$ moles per volume) is also commonly called ``ideal gas law'';
  for chemically reacting mixtures it is \emph{not} equivalent to $\pp=\Rspec\temp\dens$. Some literature objects to calling dissociation a ``non-ideal'' effect.}

Dissociation is a special case of reversible chemical reactions. 
In practice \emph{irreversible} reactions are important, 
e.g.\ in fuel-oxidizer mixtures passing through an engine. 
But there cases of non-uniqueness can be found easily, even for \emph{normal} steady shocks,
with one solution a detonation, the other solution a weaker shock that does not ignite the mixture.

In any case the results of this article confirm benign behaviour of shock polars for a much wider range of temperatures.

In section \ref{section:nonideal} we argue our results cannot hold for general (non-ideal) gas assuming only convex equation of state.

\section{Outline of the ideal polar argument}
\label{section:informal}

Since the precise argument is long, we first outline the core idea by an informal argument based on mass flux. 
  The argument gives a clear intuitive idea why critical-type shocks are generally subsonic.

\begin{figure}
  \centerline{\scriptsize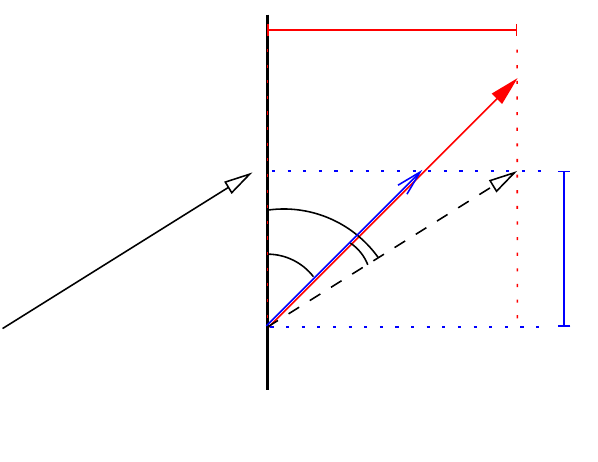}
  \caption{Across shocks, normal mass flux $\jj\dotp\nn$ and tangential velocity $\uu\dotp\ts$ are continuous
    ($\dens_0=1$ scaling in this figure for clarity)}
  \label{fig:shockben}
  \centerline{\scriptsize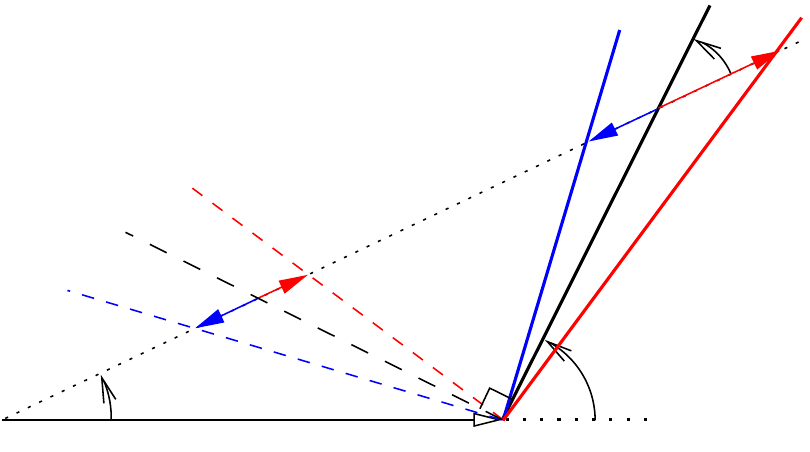}
  \caption{
    With $\turn$ and $\uu_0=\jj_0$ held fixed, 
    increasing $\bet_0$ decreases $\uu,\jj$. ($\dens_0=1$ in this figure for clarity.)
  }
  \label{fig:deltajq}
  \centerline{\scriptsize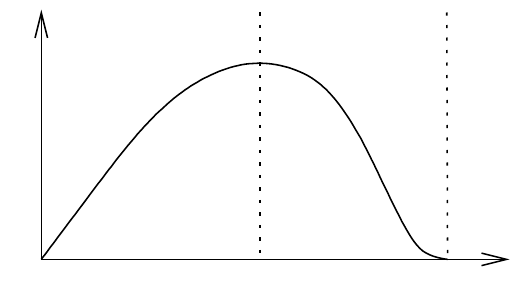}
  \caption{Mass flux as a function of speed at constant entropy and total enthalpy}
  \label{fig:jqlin}
\end{figure}

A good coordinate-free representation of the turning angle $\turn$ is to represent it as
\begin{alignat*}{5} \turn = \bet_0-\bet \end{alignat*} 
where (see fig.\ \ref{fig:shockben} and \ref{fig:deltajq}) subscript $0$ indicates upstream side, no subscript means downstream, and $\bet,\bet_0$ are the angles between shock and velocity $\uu$ or $\uu_0$;
the length of $\uu$ is $\ua$, etc.
Tangential velocity $\ut$ is continuous, as is the normal part $\jn$ of mass flux $\jj=\dens\uu$, so the formulas
\begin{alignat*}{5} \bet = \arccos\frac{\ut}{\ua} = \arcsin\frac{\jn}{\ja}\end{alignat*} 
are especially convenient since the only ``unknowns'' in $\turn$ involving thermodynamics are downstream speed $\ua$ and mass flux $\ja$, 
the rest are purely geometric from $\bet_0,\jau,\uau$. 

The upper half of the shock polar (fig.\ \ref{fig:polypolar}) is the curve of $\uu$ resulting from increasing $\bet_0$ 
from the Mach angle $\arcsin(1/\Mach_0)$ (vanishing shock) to $90^\circ$ (normal shock). 
At a critical-type shock by definition $\turn$ stagnates; 
holding $\turn$ and $\uu_0,\jj_0$ fixed while increasing $\bet_0$ (see fig.\ \ref{fig:deltajq})
decreases both $\ua$ \emph{and} $\ja$. 

Immediately a classical observation of fluid dynamics comes to mind (fig.\ \ref{fig:jqlin}): 
decreasing speed $\ua$ will\footnote{assuming constant per-mass total enthalpy $\Hpm$ (true for shocks) and entropy $\spm$ (tiny change if shock strength small)} decrease mass flux $\ja$ if and only if the flow is subsonic. 
This explains clearly why the critical-type shocks are generally on the subsonic part of the polar.

The change of mass flux with speed is quantified by the well-known formula
\begin{alignat*}{5} \frac{\dehug\ln\ja}{\dehug\ln\ua} &= 1-\Mach^2. \myeqlab{eq:oMM-djdq} \end{alignat*}
This suggests how to show uniqueness of critical-type shocks:
downstream Mach number $\Mach$ is usually \strictly decreasing\footnote{all ``decreasing'' and ``increasing'' in the ``strictly'' sense} 
from vanishing to normal. Then the right-hand side of \eqref{eq:oMM-djdq} would be \strictly increasing.
So it cannot equal more than once the left-hand side, which is \emph{de}creasing when specialized to \emph{critical-type shocks}.
This last property is less obvious and requires a detailed calculation, given around \myeqref{eq:djj-dqq}.

This is probably as simple as informal arguments can be made, 
but clearly there are many loose ends.
In the following sections we will carefully fill in every detail and impose missing assumptions.

\section{Shock relations}

\noindent 
Let $\nn$ be a unit normal to the shock, $\jmp{a}=a-a_0$ jump of some quantity $a$ from upstream to downstream, $\avg{a}=\half(a+a_0)$ average of both sides.
Conservation of mass, momentum, energy for a steady straight shock: 
\begin{alignat*}{5} 0 &= \jmp{\jn}, \myeqlab{eq:masscons}
\\ 0 &= \jmpl{\jn\uu}+\jmpl{p\nn}, \myeqlab{eq:momcons}
\\ 0 &= \jmpl{\jn(\epm+\frac{|\uu|^2}{2})}+\jmpl{\pp\un}. \myeqlab{eq:encons}\end{alignat*}
Second law of thermodynamics (entropy inequality):
\begin{alignat*}{5} 0 &\leq \jmpl{\jn\spm}. \myeqlab{eq:entropyineq}\end{alignat*} 
Tangential part of momentum conservation \myeqref{eq:momcons}: 
\begin{alignat*}{5} 0 = \jmpl{\jn\ut} \overset{\jmp\jn=0}{=} \jn \jmp{\ut} ; \end{alignat*} 
for shocks $\jn\neq 0$ so that 
\begin{alignat*}{5} \jmp{\ut}=0 ,\end{alignat*}
so $\jmp{\uu}$ is normal to the shock. 
On the other hand $\jmp{\jj}$ is \emph{tangential} to the shock, by mass conservation \myeqref{eq:masscons}.
We orient shock normals so that $\jn>0$, then the entropy inequality \myeqref{eq:entropyineq} reduces to 
\begin{alignat*}{5} 0 &\leq \jmp{\spm}. \end{alignat*} 
Normal part of momentum conservation \myeqref{eq:momcons}: 
\begin{alignat*}{5} 0 &= \jn\jmp{\supeq{\un}{=\idens\dens\un=\idens\jn}}+\jmp{\pp}   \myeqlab{eq:jjpv1}
\\&= (\jn)^2\jmp{\idens}+\jmp{\pp}  \qiq
 \frac{\jmp{\pp}}{-\jmp\idens} = (\jn)^2 . \myeqlab{eq:jjpv} \end{alignat*} 
From this we can also derive the useful
\begin{alignat*}{5} |\jmp\uu|^2 \overset{\jmp{\ut}=0}{=} (\jmp\un)^2 = \big(\jmpl{\idens\subeq{\dens\un}{=\jn}}\big)^2 = (\jn\jmp{\idens})^2 \topref{eq:jjpv}{=} -\jmp\idens\jmp{\pp}. \myeqlab{eq:upv} \end{alignat*} 
Using that $\jmp{\uu}$ is a shock normal 
we may obtain a unit normal $\nn=\jmp{\uu}/|\jmp{\uu}|$, so
\begin{alignat*}{5} - \jmp\pp \topref{eq:jjpv1}{=} \jn_0 \jmp{\un} = \dens_0\uu_0\dotp\nn \jmp\uu\dotp\nn 
=
\dens_0\uu_0\dotp\jmp\uu , \myeqlab{eq:jp-jux}\end{alignat*}
so if the coordinates are chosen to let $\uu_0$ point horizontal right, then 
\begin{alignat*}{5} \pp = \pp_0 + \dens_0 \ua_0 (\ua_0-\ux)  . \end{alignat*} 
Hence pressure can serve as a horizontal coordinate in the $\uu$ shock polar plane. 

Energy: in \myeqref{eq:encons} use $\pp\un=\pp\idens\dens\un$ to get
\begin{alignat*}{5} 0 
&= 
\jn\jmpl{\subeq{\supeq{\pp\idens+\epm}{=\hpm}+\frac{|\uu|^2}{2}}{=\Hpm}} 
\qiq
    \jmp{\Hpm} = 0,  \myeqlab{eq:bern}
\end{alignat*} 
with $\Hpm$ total enthalpy per mass. The shock relations can be reduced further to a single scalar relation: applying $\jmpl{a^2}=2\avg a\jmp{a}$ to $a=\idens$,
\begin{alignat*}{5} 0 
&\topref{eq:bern}{\underset{\jmp{\ut}=0}{=}} 
\jmp{\hpm}+\jmpl{\frac{\idens^2(\dens\un)^2}{2}}    
= \jmp{\hpm}+(\jn)^2\jmpl{\frac{\idens^2}{2}}   \myeqlab{eq:hhug1}
\\&= \jmp{\hpm} + (\jn)^2 \avg\idens \jmp{\idens} 
\topref{eq:jjpv}{=} \jmp{\hpm} - \avg\idens \jmp{\pp}.  \myeqlab{eq:hhug}\end{alignat*} 
This \defm{Hugoniot relation} involves no (macroscopic) velocities, only thermodynamic quantities.
Using $\hpm=\epm+\pp\idens$ and applying $\jmpl{ab}=\avg a\jmp b+\avg b\jmp a$ to $a=\pp$, $b=\idens$ shows
\begin{alignat*}{5} 0 = \jmp{\epm} + \avg\pp\jmp{\idens},  \myeqlab{eq:ehug} \end{alignat*} 
a common alternate form.

\section{Ideal polar}
\label{section:idealpolar}

For the rest of the article we choose units so that $\Rspec=1$, to simplify calculations. 
The resulting ideal gas law $\idens=\stemp/\pp$ can be used in the Hugoniot relation \eqref{eq:ehug}:
\begin{alignat*}{5} 0 = 2\jmp{\epm} + (\pp+\pp_0)\jmp{\frac{\stemp}{\pp}}; \end{alignat*}
this is essentially a \emph{quadratic} equation for $\pp$, with solution
\begin{alignat*}{5} \frac{\pp}{\pp_0}
= \frac{\jmpl{\epm+\stemp/2}}{\stemp_0} + \sqrt{\big(\frac{\jmpl{\epm+\stemp/2}}{\stemp_0}\big)^2+\frac{\stemp}{\stemp_0}} , \myeqlab{eq:pt-formula} \end{alignat*}
where we may ignore the $-\sqrt{\cdot}$ second solution (assuming positive temperatures and pressures). 
From this we also see
\begin{alignat*}{5} \frac{\idens_0}{\idens} = \frac{\pp}{\pp_0} \frac{\stemp_0}{\stemp} 
= \frac{\jmpl{\epm+\stemp/2}}{\stemp} + \sqrt{\big(\frac{\jmpl{\epm+\stemp/2}}{\stemp}\big)^2+\frac{\stemp_0}{\stemp}} .
\myeqlab{eq:vt-formula}
\end{alignat*}

As noted in \eqref{eq:jp-jux} there is a linear relationship between $\jmp\pp$ and $\jmp\ux$, if we rotate $\uu_0$ to be horizontal pointing right, 
and having obtained $\pp$ and $\idens$ we can use \eqref{eq:upv} $|\jmp\uu|^2=\jmp\pp\jmpl{-\idens}$
and then 
\begin{alignat*}{5} \uy=\jmp\uy=\pm\sqrt{|\jmp\uu|^2-|\jmp{\ux}|^2} \myeqlab{eq:uy-formula} \end{alignat*}
to calculate all of $\uu$. 
So the steps from \eqref{eq:pt-formula} provide a semi-explicit formula for the shock polar, 
with $\stemp$ as parameter instead of $\ux$. 
Unlike the polytropic case there is no explicit formula for $\stemp$ as function of $\ux$ except perhaps for special $\epm=\epm(\stemp)$.

Although some of our results can be obtained directly from \eqref{eq:pt-formula} etc.,
it would be cumbersome since taking derivatives quickly produces unwieldy expressions 
whose components do not have obvious physical meaning. 
But some quick insights are already possible.

Fig.\ \ref{fig:idealdent} shows an example with an artificial $\epm(\stemp)$ 
so that $2\epm/\Rspec\stemp$ increases from 3 to 5 over some temperature interval. 
The corresponding shock polar is non-convex, 
with dent sufficiently strong that some velocity turning angles $\turn$ permit four rather than two solutions. 
By adding more intervals of increase more dents could be created, allowing any number of solutions. 

In physical fluids the increase is generally not as sharp as in fig.\ \ref{fig:idealdent} left (cf.\ fig.\ \ref{fig:foh}). 
We will find that for ideal gas the increase that permits more than two solutions 
is characterized precisely by the familiar condition of convex equation of state.

\section{Assumptions}
\label{section:assumptions}

\subsection{General gas}

We consider fluid in thermal equilibrium described by an equation of state
\begin{alignat*}{5} \epm = \epm(\idens,\spm) \end{alignat*} 
with $\epm$ internal energy per mass, $\idens=1/\dens$ volume per mass, $\spm$ entropy per mass. 
We assume $\epm$ is differentiable as often as needed for our purposes.\footnote{There is little benefit in burdening the discussion with derivative counting, because usually $\epm$ is very smooth away from phase transitions, rather non-smooth across them.} As always
\begin{alignat*}{5} d\epm = \stemp d\spm - \pp d\idens ,  \myeqlab{eq:de}\end{alignat*} 
where $\pp$ is pressure, $\stemp$ temperature, or equivalently
\begin{alignat*}{5} d\hpm = \stemp d\spm + \idens d\pp  \myeqlab{eq:dh} \end{alignat*} 
for enthalpy per mass $\hpm=\epm+\pp\idens$. 

$\epm$ has ``standard'' coordinates $\idens,\spm$, for $\hpm$ they are $\pp,\spm$. 
Subscript notation $\epm_\spm$ indicates partial derivatives with the other standard coordinate held fixed
(e.g.\ $\dat\epm\spm\idens$, not $\dat\epm\spm\pp$). 

We make the following assumptions: 
\begin{enumerate} 
\item $\stemp=\epm_\spm>0$ and $\pp=-\epm_\idens>0$.
\item Thermodynamic stability: 
  the matrix $\epm''$ of second derivatives is positive definite. 
  That is equivalent\footnote{see e.g.\ \cite[p.\ 64]{landau-lifshitz-thermodynamics} for details} to the three conditions
  \begin{alignat*}{5} \csnd^2 = \dat\pp\dens\spm = -\idens^2 \dat\pp\idens\spm = \idens^2 \epm^{}_{\idens\idens} > 0, \myeqlab{eq:cc}\end{alignat*} 
  so that sound speed $\csnd$ is well-defined and positive, and 
  \begin{alignat*}{5} \cvspec = \dat\epm\stemp\idens = \dat\epm\spm\idens / \dat\stemp\spm\idens = \epm^{}_\spm / \epm^{}_{\spm\spm} > 0, \end{alignat*} 
  and finally $\cpspec=\dat\hpm\stemp\pp>\cvspec$.
\item Convex equation of state: 
  \begin{alignat*}{5} \pddat\pp\idens\spm > 0 \myeqlab{eq:gennon-pvv}\end{alignat*} 
  (the left-hand side divided by $2\dens^3\csnd^2$
  is sometimes called fundamental derivative). 
  Equivalent forms (see also \cite{thompson-pof1971}):
  via $\dat\pp\idens\spm=-\dens^2\csnd^2$
  \begin{alignat*}{5} \dat{\ln(\dens\csnd)}{\ln\dens}\spm > 0, \myeqlab{eq:dlnrhoc-dlnr-gennon}\end{alignat*} 
  often also written
  \begin{alignat*}{5} \dat{\ln\csnd}{\ln\dens}\spm > -1. \myeqlab{eq:dlncdlnr-gennon}\end{alignat*}
  Using $\dat{\pp}{\idens}{\spm}=-(\dens\csnd)^2<0$, \myeqref{eq:gennon-pvv} is equivalent to
  \begin{alignat*}{5} \pddat\idens\pp\spm > 0 \myeqlab{eq:gennon-vpp}\end{alignat*}
  (if $\pp$ is a convex decreasing function of $\idens$, then $\idens$ is a convex decreasing function of $\pp$, which is also easy to represent graphically).
  Using $\idens=\hpm_\pp$ 
  we obtain the last equivalent form
  \begin{alignat*}{5} \hpm_{\pp\pp\pp}
  > 0.  \myeqlab{eq:hppp-gennon} \end{alignat*} 
\end{enumerate}

To our knowledge any fluid fairly described as ``ideal gas'' also satisfies the assumptions above.
Some assumptions may be violated by a few examples of more or less realistic \emph{non-ideal} materials, at certain temperatures and pressures;
see \cite{ivanov-novikov-1961,lambrakis-thompson-1972,zamfirescu-guardone-colonna-jfm2008}
regarding non-convex equations of state.

\subsection{Ideal gas}
\label{section:ideal}

We specialize the assumptions to ideal gas, $\stemp=\pp\idens$. First,
\begin{alignat*}{5} \epm_{\idens} &= -\pp = -\stemp/\idens \csep \epm_{\spm} = \stemp \qiq 0=\idens \epm_{\idens}+\epm_{\spm}. \end{alignat*}
This PDE is easily solved by the method of characteristics:
\begin{alignat*}{5} \epm &= \hat\epm(\spm-\ln\idens)  \qiq  \stemp = \epm_{\spm} = \hat\epm'(\spm-\ln\idens). \myeqlab{eq:tes}\end{alignat*}
for an arbitrary function $\hat\epm$.
By assumption $\hat\epm'=\stemp>0$, and besides $\epm_{\spm\spm}>0$ shows $\hat\epm''>0$, 
so $\hat\epm$ and $\hat\epm'$ are both \strictly increasing and therefore invertible. 
We may write either $\epm=\epm(\temp)$ or $\temp=\temp(\epm)$.
This is Joule's second law that energy is a function of temperature alone.\footnote{It is often used as definition of ideal gas, 
but Joule's law is strictly weaker because it also holds for the van der Waals equation $(\pp+a/\idens^2)(\idens-b)=\temp$ in the $a=0$ but $b>0$ special case.}
For functions of $\stemp$ alone, such as $\epm$, we may use subscript notation $\epm^{}_\stemp$ without ambiguity.

\newcommand{\tmin}{\stemp_{\min}}
\newcommand{\tmax}{\stemp_{\max}}
We assume $\epm$ is defined in some interval $(\tmin,\tmax)$, 
and differentiable as often as we need\footnote{non-smoothness mostly occurs near phase transitions where gas is far from ideal}. 

For ideal gas entropy has a semi-explicit formula: 
\begin{alignat*}{5} 
d\epm \topref{eq:tes}{=} \subeq{\hat\epm'(\spm-\ln\idens)}{=\stemp}d(\spm-\ln\idens) \qiq
\spm = \ln\idens + \int \frac1{\stemp(\epm)} d\epm , \myeqlab{eq:ideal-s}\end{alignat*}
or analogously for $\hpm$ replacing $\epm$, 
\begin{alignat*}{5} \spm = - \ln \pp + \int \frac1{\stemp} d\hpm. \myeqlab{eq:ideal-s-h}\end{alignat*} 
Note $\hpm=\epm+\pp\idens=\epm+\stemp$, so $\hpm$ is also a function of temperature alone; $\hpm_\stemp=\epm_\stemp+1>1>0$, 
so it may also be used as coordinate. 

Now simpler equivalent conditions to our assumptions can be derived:
$\cvspec=\epm_\stemp$, $\cpspec=\hpm_\stemp=\epm_\stemp+1$ and
\begin{alignat*}{5} \csnd^2 = \dat\pp\dens\spm = \dat\pp\dens\epm + \dat\pp\epm\dens \dat\epm\dens\spm \overset{\pp=\stemp\dens}{\underset{\stemp=\stemp(\epm)}{=}} 
\stemp + \stemp_\epm\rho \frac{\pp}{\rho^2} = \stemp(\stemp_\epm+1), \myeqlab{eq:ccte} \end{alignat*}
so $\stemp_\epm>0$ already forces $\epm''$ positive definite; no additional assumptions are required. 

After some calculation using \myeqref{eq:ccte}, convex equation of state \myeqref{eq:dlnrhoc-dlnr-gennon} is seen to be equivalent to 
\begin{alignat*}{5} 
\frac{\stemp\epm^{}_{\stemp\stemp}}{\epm^{}_\stemp} &< (1+\epm^{}_\stemp)(1+2\epm^{}_\stemp). \myeqlab{eq:ett-gennon}
\end{alignat*}
This is an upper bound on the growth of heat capacity $\epm_\stemp$ as temperature rises.

Finally we note that our assumptions, in particular the ideal gas law, are only needed in a neighbourhood of the Hugoniot curve associated with the shock polar at hand.

\section{Monotonicities for ideal normal shocks}

Constructing a set of solutions $(\pp,\spm)$ of the Hugoniot relation
generally requires additional assumptions and long discussion for non-ideal gas \cite{bethe,weyl-shock-waves,menikoff-plohr}.
Proving the set is a curve without disconnected additional subsets of ``exotic'' shocks is not trivial,
with inaccuracies in past work pointed out by \cite{henderson-menikoff}. 
But in the ideal case we already have a Hugoniot curve in the form \eqref{eq:pt-formula}:

\begin{alignat*}{5} \frac{\pp}{\pp_0} \topref{eq:pt-formula}{=} \frac{\jmpl{\epm+\stemp/2}}{\stemp_0} + \sqrt{\big(\frac{\jmpl{\epm+\stemp/2}}{\stemp_0}\big)^2+\frac{\stemp}{\stemp_0}} . \end{alignat*}
\newcommand{\pmax}{\pp_{\max}}
\newcommand{\pmin}{\pp_{\min}}
$\epm_\stemp>0$, so $\pp$ is clearly \strictly increasing in $\stemp\in(\tmin,\tmax)$, with value range $(\pmin,\pmax)$. 
Now that we have assumptions, monotonicity for some other variables is also true.

Focus on the $\jmp\stemp>0$ ``forward'' branch of the Hugoniot curve. 
Since $\epm_\stemp>0$, necessarily $\jmp\epm>0$, and by the $\epm$ form \eqref{eq:ehug} of the Hugoniot relation
that requires 
\begin{alignat*}{5} \jmp\idens<0 . \myeqlab{eq:idensneg}\end{alignat*}

%\begin{figure}
%  \centerline{\fbox{\input{pshug.pdf_t}}}
%  \caption{Left: $\hug_\spm>0$ means the Hugoniot $\spm=S(\pp)$ cannot become vertical. Right: $S_\pp=0$ is not possible.}
%\end{figure}

We need to show that $\spm$ is \strictly increasing. 
To simplify notation use $\pp,\spm$ as standard coordinates for $\hug$, same as for $\hpm$, and use subscripts to denote partial derivatives.
We parametrize the Hugoniot curve as $\spm=S(\pp)$.
Consider its defining equation
\begin{alignat*}{5}
  0 = \hug(\pp,S(\pp)) \topref{eq:hhug}{=} \hpm - \hpm_0 - \frac{\idens + \idens_0}{2} (\pp-\pp_0) .     \myeqlab{eq:Khug}
\end{alignat*}
Taking $d/d\pp$ derivatives we find by the multidimensional chain rule that
\begin{alignat*}{5}
  0 &= \hug_\pp(\pp,S(\pp)) + \hug_\spm(\pp,S(\pp)) S_\pp(\pp). \myeqlab{eq:Kp-Sp}
\end{alignat*}
Another derivative shows (with $(\pp,S(\pp))$ arguments omitted)
\begin{alignat*}{5}
  0 &= \hug_{\pp\pp} + (2\hug_{\pp\spm}+\hug_{\spm\spm}S_\pp) S_\pp + \hug_\spm S_{\pp\pp}. \myeqlab{eq:Kpp-Spp}
\end{alignat*}
A third derivative yields (with ``...'' coefficients that will not matter)
\begin{alignat*}{5}
  0 &= \hug_{\pp\pp\pp} + ... S_\pp + ... S_{\pp\pp} + \hug_\spm S_{\pp\pp\pp}. \myeqlab{eq:Kppp-Sppp}
\end{alignat*}

Now we calculate $\hug$ derivatives, starting with $\hug_\spm$: 
\begin{alignat*}{5} 
  \hug_\spm = \hpm_\spm + ... \jmp\pp \overset{\pp=\pp_0}{=} \temp
\end{alignat*}
which is positive. In the \emph{ideal} case $\hug_\spm$ is in fact positive for \emph{any} $\pp,\spm$:
\begin{alignat*}{5} 
  \dat\hug\spm\pp = 
  \dat\hug\stemp\pp ~\dat\stemp\spm\pp 
  &\underset{\idens=\stemp/\pp}{\topref{eq:Khug}{=}} 
  (\hpm_\stemp - \frac{\jmp\pp}{2\pp}) / \dat\spm\stemp\pp           
  \\&\topref{eq:ideal-s-h}{=}
  (\subeq{\hpm_\stemp}{>1}-\half+\frac{\pp_0}{2\pp}) / (\frac{\hpm_\stemp}{\stemp})
  > 0 . \myeqlab{eq:Ksp-positive}\end{alignat*}
%\uimp{should just DRAW diagrams, to make this more readable.})

Now consider $\hug_\pp$, which is more delicate:
\begin{alignat*}{5} 
%\dat\hug\pp\spm 
\hug_\pp
&\topref{eq:Khug}{=} \subeq{\hpm_{\pp}}{=\idens} - \avg\idens - \half \subeq{\dat\idens\pp\spm}{=-1/(\dens\csnd)^2} \jmp\pp       \myeqlab{eq:Kpone}
\\&= \half \jmp\idens + \frac1{2(\dens\csnd)^2} \subeq{\jmp\pp}{=-(\jn)^2\jmp\idens} 
\\&\overset{\Machn=\un/\csnd}{=} \half \big( 1 - (\Machn)^2\big) \subeq{\jmp\idens}{<0}.
\myeqlab{eq:Hpats} \end{alignat*}
This is zero at vanishing, i.e.\ $\pp=\pp_0$ and $\spm=\spm_0$, so that $\idens=\idens_0$; hence by \myeqref{eq:Kp-Sp} $S_\pp=0$ there as well. 
So we take another derivative:
\begin{alignat*}{5} \hug_{\pp\pp} &\topref{eq:Kpone}{=} \subeq{\hpm_{\pp\pp}-2\frac{\dat\idens\pp\spm}{2} }{=0}-\half \pddat\idens\pp\spm \jmp\pp  = - \half \subeq{\hpm_{\pp\pp\pp}}{>0}\jmp\pp , \myeqlab{eq:Hppats}\end{alignat*}
which is also zero at vanishing, so that by \myeqref{eq:Kpp-Spp} $S_{pp}=0$ there. On the other hand \myeqref{eq:Hppats} also shows $\hug_{\pp\pp}$ is negative for any $\pp>\pp_0$, regardless of $\spm$, due to convex eos $\hpm_{\pp\pp\pp}>0$ \myeqref{eq:hppp-gennon}. Accordingly
\begin{alignat*}{5} \hug_{\pp\pp\pp} &\topref{eq:Hppats}{=} - \half \subeq{\hpm_{\pp\pp\pp}}{>0} - \hpm_{\pp\pp\pp\pp}\jmp\pp \end{alignat*}
is negative at vanishing, where we have already shown $S_{\pp}=S_{\pp\pp}=0$, so \myeqref{eq:Kppp-Sppp} shows $S_{\pp\pp\pp}>0$ there. 
So we find $S_{\pp}>0$ for $\pp$ near to but larger than $\pp_0$.

Assume that $S_{\pp}$ returns to $0$ at some larger $\pp_1$, which we can take \emph{minimal}, so that $S_{\pp}$ returns to $0$ \emph{from above},
meaning $S_{\pp\pp}\leq 0$ in $\pp_1$. But this contradicts \myeqref{eq:Kpp-Spp} which shows
\[ S_{\pp\pp} = - \subeq{ \frac1{\hug_{\spm}} }{>0} ( ... \subeq{S_{\pp}}{=0} + \subeq{\hug_{\pp\pp}}{<0} ) > 0 \quad\text{at $\pp_1$,} \] 
where we use that $S_{\pp}=0$ at $\pp_1$ and that we have already shown $\hug_{\pp\pp}<0$ and $\hug_{\spm}>0$ for any $\pp>\pp_0$. So such a $\pp_1$ cannot exist.
Therefore $S_{\pp}>0$ for all $\pp>\pp_0$:
\begin{alignat*}{5}
  \text{$\spm$ is \strictly increasing on the entire forward Hugoniot.}
\end{alignat*}
$0<S_{\pp}\topref{eq:Kp-Sp}{=}-\hug_\pp/\hug_\spm$ with $\hug_\spm>0$ shows $\hug_{\pp}<0$, so \myeqref{eq:Hpats} shows $\Machn<1$ along the forward Hugoniot --- as expected.

Analogously we find $\Machn>1$ along the backward Hugoniot curve.
Since the Hugoniot relation is symmetric under exchanging the downstream and upstream state,
$\Machn>1$ on backward curves means $\Machn_0>1$ on forward curves. So: 
\begin{alignat*}{5} \text{Normal velocities are subsonic downstream and supersonic upstream.} \myeqlab{eq:norsub}\end{alignat*}

Henceforth let $\dehug$ denote infinitesimal change along the Hugoniot curve $\{K=0\}$. 
We need a formula relating $\delta\spm$ to $\delta\jn$: consider
\begin{alignat*}{5} 0 &\topref{eq:jjpv}{=} \jmp\pp + (\jn)^2\jmp\idens , 
\\ 0 &\topref{eq:hhug1}{=} \jmp\hpm + (\jn)^2\jmpl{\frac{\idens^2}{2}} . 
\end{alignat*}
Differentials $\dehug$ along the Hugoniot curve:
\begin{alignat*}{5} 0 &= \dehug\pp &+& \dehug\big((\jn)^2\big)\jmp{\idens} &+& (\jn)^2\dehug\idens ,
\\ 0 &= \subeq{\dehug\hpm}{=\idens\dehug\pp+\stemp\dehug\spm} &+& \dehug\big((\jn)^2\big)\subeq{\jmp{\frac{\idens^2}{2}}}{=\avg\idens\jmp\idens} &+& (\jn)^2\idens \dehug\idens . \end{alignat*}
First times $\idens$ subtracted from second:
\begin{alignat*}{5} 0 &= \stemp \dehug\spm - \half (\jmp\idens)^2  \dehug((\jn)^2) \\
\qiq \dehug\spm &= \frac{(\jmp\idens)^2}{\stemp}\jn\dehug\jn 
= \frac{-\jmp\pp\jmp\idens}{\pp\idens}\frac{\dehug\jn}{\jn}. \myeqlab{eq:dsdaa}\end{alignat*} 
We have shown $\pp$ is increasing, so $\jmp\pp>0$, while $\jmp\idens<0$ \myeqref{eq:idensneg}, so \myeqref{eq:dsdaa} shows
\begin{alignat*}{5} \sign \dehug\spm = \sign \dehug\jn (= \sign\dehug\jn_0)\end{alignat*} 
regardless of shock strength. 

Finally $\dens_0\un_0=\jn_0$ means $\sign\dehug\un_0=\sign\dehug\jn_0$, and then $\Machn_0=\un_0/\csnd_0$ shows $\sign\dehug\Machn_0=\sign\dehug\un_0$.

All combined: along Hugoniot curves for ideal convex eos, 
\begin{alignat*}{5} \text{$\epm,\hpm,\dens\un,\stemp,\spm,\pp,\un_0,\Machn_0$ are \strictly increasing} \myeqlab{eq:genmon}\end{alignat*} 
(under the assumptions in section \ref{section:assumptions}).

Many other variables are \emph{not} monotone. The calculation for $\dens\csnd$ is also needed for our main result, so we treat it first.

  Note $1/(\dens\csnd)^2=-\hpm_{\pp\pp}$; we use $\hpm_\spm=\stemp=\pp\idens=\pp\hpm_\pp$ to calculate 
  \begin{alignat*}{5} \hpm_{\pp\pp\spm} = (\pp\hpm_\pp)_{\pp\pp} = \pp\hpm_{\pp\pp\pp} + 2\hpm_{\pp\pp} \end{alignat*} 
  so that 
  \begin{alignat*}{5} \dehug(\hpm_{\pp\pp}) 
  &= \hpm_{\pp\pp\pp}\dehug\pp + \hpm_{\pp\pp\spm}\dehug\spm 
  = \hpm_{\pp\pp\pp}(\dehug\pp+\pp\dehug\spm) + 2\hpm_{\pp\pp}\dehug\spm. \myeqlab{eq:delhpp} \end{alignat*}
  Although $\hpm_{\pp\pp\pp}>0$, the last term has the wrong sign due to $\hpm_{\pp\pp}=-1/(\dens\csnd)^2<0$.
We can construct eos where $\hpm_{\pp\pp\pp}$ suddenly drops to near-zero, by choosing an $\epm_{\temp\temp}$ near the limit allowed by the inequality \myeqref{eq:ett-gennon}
  equivalent to convex eos, without much initial change to the up to second derivatives of $\hpm$ determining $\dehug\spm/\dehug\pp$
  (see \myeqref{eq:Kp-Sp}, \myeqref{eq:Ksp-positive} and \myeqref{eq:Hpats}). Then the right-hand side of \myeqref{eq:delhpp} is negative.
  So $\dens\csnd$ can sometimes decrease, although it ``usually'' increases.

  $\Machn=1$ in the vanishing limit, but $\Machn<1$ for admissible shocks under our assumptions, so clearly $\Machn$ is decreasing somewhere.
  But 
  \begin{alignat*}{5} \dens\csnd = \frac{\dens\un}{\Machn} \end{alignat*}
  and we have already argued that mass flux $\dens\un$ is increasing, so if $\Machn$ was \emph{always} decreasing, then $\dens\csnd$ would always increase
  --- we have shown it does not. So $\Machn$ need not be monotone. 

  For sound speed $\csnd$ itself we note that in the limit of vanishing shocks
  \begin{alignat*}{5} \frac{\dehug\csnd}{\dehug\pp}
    = \dat{\csnd}{\pp}{\spm} + \dat{\csnd}{\spm}{\pp} \frac{\dehug\spm}{\dehug\pp}
    = \dat{\csnd}{\dens}{\spm} / \subeq{\dat{\pp}{\dens}{\spm}}{=\csnd^2>0} + \dat{\csnd}{\spm}{\pp} \subeq{\frac{\dehug\spm}{\dehug\pp}}{\conv 0} , \end{alignat*}
  but \myeqref{eq:dlnrhoc-dlnr-gennon} permits either sign of $\dati{\csnd}{\dens}{\spm}$, so that $\csnd$ can be increasing or decreasing even for weak shocks.

  To discuss $\idens$ we consider compression ratio $\idens_0/\idens$.
  For polytropic eos, i.e.\ $\epm_{\stemp}=\fod/2$ constant, it is well-known that $\idens$ converges to a constant in the limit of infinitely
  strong shocks. For ideal gas, if $\fod$ is constant on a sufficiently wide $\stemp$ interval, so that
  $\epm/\stemp\rightarrow\fod/2$, and $\stemp_0/\stemp\rightarrow 0$ as well as $\epm_0/\epm\rightarrow 0$, then
  \myeqref{eq:vt-formula} shows
  \begin{alignat*}{5} \frac{\idens_0}{\idens} \rightarrow \frac{\fod+1}{2} + \sqrt{ (\frac{\fod+1}{2})^2 + 0 } = \fod+1 , \end{alignat*}
  which is familiar. However, in the ideal case we are flexible to use several wide intervals, changing $\fod$ between them.
  The convex eos \myeqref{eq:ett-gennon} permits decrease and increase of $\epm_\temp$, 
  limiting only the rate of the latter; thermodynamic stability only requires $\epm_\temp>0$.
  So we can choose some non-monotone sequence of $\fod+1$, making intervals wide enough to let $\idens_0/\idens$ come arbitrarily close to each value so that it cannot be monotone either.

  Finally, normal downstream velocity $\un=\jn\idens$ is not monotone even for (say) $\gisen=7/5$ polytropic, as some simple explicit calculations show; 
  $\un$ is decreasing for weak but increasing for strong shocks.

  \section{Monotonicities along the shock polar}

For the shock polar upstream $\idens_0,\spm_0$ are fixed, so the Hugoniot curve determines $\idens,\spm$ along the polar.
The Hugoniot also determines $\un_0=\idens_0\jn_0$ by $(\jn_0)^2=-\jmp\pp/\jmp\idens$. 
The Hugoniot only describes normal shocks, but for the polar $\uau$ is fixed, and so is $\ut=\ut_0=\pm\sqrt{\uau^2-(\un_0)^2}$ (choosing the sign chooses one half of the polar). 

For normal shocks 
\begin{alignat*}{5} \hpm + \half (\un)^2 \topref{eq:bern}{\underset{\ut=\ut_0}{=}} \hpm_0 + \half (\un_0)^2 \end{alignat*} 
does not necessarily yield monotone $\un$, since $\un_0$ also varies. In contrast $\uau$ is fixed for the polar
so that 
\begin{alignat*}{5} \hpm + \half \ua^2 = \hpm_0 + \half \uau ^2 \end{alignat*} 
with $\hpm$ \strictly increasing (from vanishing to normal) shows:
\begin{alignat*}{5} \text{$\ua$ is \strictly decreasing.} \myeqlab{eq:udecr} \end{alignat*}

We also need mass flux $\ja$ monotonicity. 
First we reprove the classical result on mass flux change with speed:
\begin{alignat*}{5} d\Hpm - \ua d\ua = d\hpm & , \\
d\hpm &= \stemp d\spm + \idens & d\pp & , \\
&& d\pp &=& \subeq{\dat\pp\dens\spm}{=\csnd^2} d\dens + \dat\pp\spm\dens d\spm ,\end{alignat*} 
so setting $d\Hpm=0=d\spm$ we find 
\begin{alignat*}{5} \frac1{\dens} \dat\dens\ua{\spm,\Hpm} = -\csnd^{-2}\ua , \end{alignat*}
so that we obtain the familiar
\begin{alignat*}{5} \dat{(\dens\ua)}{\ua}{\spm,\Hpm} = \dens(1-\Mach^2) . \myeqlab{eq:rasH}\end{alignat*} 

Now for shocks total enthalpy $\Hpm$ is still constant \eqref{eq:bern}; although entropy $\spm$ is not, we have shown it is \strictly increasing along the Hugoniot curve. This allows modifying the argument to
\begin{alignat*}{5} \dehug(\dens\ua) 
&= \dat{(\dens\ua)}\ua{\spm,\Hpm} \dehug\ua + \dat{(\dens\ua)}\spm{\ua,\Hpm} \dehug\spm 
\\&\topref{eq:rasH}{=} \dens (1-\Mach^2) \dehug\ua + \ua \dat\dens\spm\hpm \dehug\spm . \myeqlab{eq:didadu}\end{alignat*} 
Last term: for \emph{ideal} gas $0=d\hpm$ means, by $\hpm=\hpm(\stemp)$ with $\hpm_\stemp>1$, that $0=d\stemp$, 
so with \eqref{eq:ideal-s} $ds=d\ln\idens+\epm_\stemp d\stemp/\stemp$ we find
\begin{alignat*}{5} \dat\dens\spm\hpm = -\dens. \end{alignat*}
Substitute that into \myeqref{eq:didadu} and solve for
\begin{alignat*}{5} 1-\Mach^2 = \frac{\dehug\ln\ja + \dehug\spm}{\dehug\ln\ua} . \myeqlab{eq:omMMjs}\end{alignat*} 
Consider $\Mach\leq1$; $\spm$ is \strictly increasing while $\ua$ is \strictly decreasing, so:
\begin{alignat*}{5}
\text{Mass flux $\ja$ is \strictly decreasing along the \emph{subsonic-sonic} part of the polar} \myeqlab{eq:j-decr-sub}
\end{alignat*}
(from vanishing to normal).\footnote{On the supersonic part $\ja$ may decrease or increase.}

As we discussed earlier $\csnd$ or $\dens\csnd$ or $\Machn$ are not necessarily monotone. 
But fortunately the \emph{total} Mach number $\Mach$ is \strictly decreasing once it is $\leq 1$, i.e.\ on \emph{sonic-subsonic} segments of the polar,
as we show now. 
Starting from
\begin{alignat*}{5} 1-\Mach^2 = 1-\frac{(\dens\ua)^2}{(\dens\csnd)^2} = 1 + \hpm_{pp}\ja^2 \myeqlab{eq:oMjc} \end{alignat*}
calculate
\begin{alignat*}{5} \dehug(1-\Mach^2) 
&=
\ja^2 \big( \dehug(\hpm_{\pp\pp}) + 2\hpm_{\pp\pp} \frac{\dehug\ja}{\ja} \big)
\\&\topref{eq:delhpp}{=} \ja^2 \Big( \hpm_{\pp\pp\pp}(\dehug\pp+\pp\dehug\spm) + 2\hpm_{\pp\pp} ( \dehug\spm + \frac{\dehug\ja}{\ja} ) \Big). \end{alignat*}
But for the last parenthesis we can use \eqref{eq:omMMjs}:
\begin{alignat*}{5} \dehug(1-\Mach^2) 
&=
\subeq{\ja^2}{>0} 
\Big( \subeq{\hpm_{\pp\pp\pp}}{>0}(\dehug\pp+\subeq{\pp}{>0}\dehug\spm) 
+ 2\subeq{\hpm_{\pp\pp}}{<0} (1-\Mach^2) \frac{\dehug\ua}{\ua} \Big). \myeqlab{eq:doMM}\end{alignat*}
We already know $\pp$ and $\spm$ are \strictly increasing, $\ua$ is \strictly decreasing, so whenever $1-\Mach^2$ is nonnegative \myeqref{eq:doMM} shows it is also \strictly increasing. So:
\begin{alignat*}{5} \text{$\Mach$ is \strictly decreasing on subsonic-sonic segments of the polar.} \myeqlab{eq:Mach-decr-sub}\end{alignat*}
At the vanishing limit $\Mach\approx\Mach_0>1$, but at the normal shock $\Mach=\Machn<1$ \myeqref{eq:norsub}, so there must be some point in between where $\Mach=1$. 
Take the first such point from vanishing; there cannot be another one as we just found $\Mach$ is decreasing from there onward, so:
\begin{alignat*}{5} \text{Each polar half splits into \emph{one} subsonic and \emph{one} supersonic segment.} \myeqlab{eq:sub-super-seg}\end{alignat*}
Thus sonic points are unique.

\section{Mass flux argument details}
\label{section:massfluxfinish}

\noindent Returning to section \ref{section:informal} we can finally make the details of the argument precise. 
\begin{alignat*}{5} \bet = \arcsin\frac{\jn}{\ja} \csep \jn=\jn_0=\jau\sin\bet_0 ; \end{alignat*} 
as $\bet_0$ increases $\jn$ also increases and by \myeqref{eq:j-decr-sub} $\ja$ is \strictly decreasing if $\Mach<1$. So:
\begin{alignat*}{5} \text{  $\bet$ is \strictly increasing on the \emph{subsonic-sonic} part of the upper half polar} \myeqlab{eq:bet-incr-sub}\end{alignat*}
(in the direction towards the normal shock). 
Like $\bet_0$ it increases from Mach angle $\arcsin(1/\Mach_0)$ at vanishing to $90^\circ$ at normal shocks, but of course in between values may differ.

Formulas for $\dehug\ja/\dehug\ua$ can be obtained in generality, but we only need them at a \emph{critical-type} shock where $\dehug\turn=0$,
i.e.\ $\dehug\bet=\dehug\bet_0$ (cf.\ fig.\ \ref{fig:shockben} and \ref{fig:deltajq}).
\begin{alignat*}{5} \jn = \ja \sin\bet  
\qiq  \dehug\ln\jn = \dehug\ln\ja + \cot\bet \dehug\bet \myeqlab{eq:dlnjn}\end{alignat*} 
and similarly
\begin{alignat*}{5} \dehug\ln\subeq{\jn_0}{=\jn} = \subeq{\dehug\ln\jau}{=0} + \cot\bet_0 \subeq{\dehug\bet_0}{=\dehug\bet} ; \myeqlab{eq:dlnjn0}\end{alignat*}
last minus second last equation:
\begin{alignat*}{5} (\cot\bet_0-\cot\bet) \dehug\bet_0 \creq \dehug\ln\ja . \myeqlab{eq:dlnja-dlnbet0}\end{alignat*} 
Similarly $\ua\cos\bet=\ut=\ut_0=\uau\cos\beta_0$, yielding
\begin{alignat*}{5} (\tan\bet-\tan\bet_0) \dehug\bet_0 \creq \dehug\ln\ua . \myeqlab{eq:dlnua-dlnbet0}\end{alignat*} 
Divide \myeqref{eq:dlnja-dlnbet0} by \myeqref{eq:dlnua-dlnbet0}:
\begin{alignat*}{5} \frac{\dehug\ln\ja}{\dehug\ln\ua} \creq \cot\bet_0 \cot\bet . \myeqlab{eq:djj-dqq}\end{alignat*} 

Besides, using \eqref{eq:dlnjn0} to eliminate $\dehug\bet$ in \eqref{eq:dlnjn} yields
\begin{alignat*}{5} \dehug\ln\ja \creq \dehug\ln\jn ( 1 - \frac{\cot\bet}{\cot\bet_0} ) . \end{alignat*} 
Divide $\ut=\ua\cos\bet$ by $\un=\ua\sin\bet$ to see $\cot\bet=\ut/\un$,
and likewise $\cot\bet_0=\ut_0/\un_0$,
then use $\ut=\ut_0$ to get $\cot\bet/\cot\bet_0=\un_0/\un=\idens_0/\idens$, so
\begin{alignat*}{5} \dehug\ln\ja \creq \dehug\ln\jn \frac{\jmp\idens}{\idens} .\end{alignat*} 
This we use in
\begin{alignat*}{5} \dehug\spm 
\topref{eq:dsdaa}{=} 
- \frac{\jmp\pp}{\pp}~ \frac{\jmp\idens}{\idens} \dehug\ln\jn
\creq 
-\frac{\jmp p}{p} \dehug\ln\ja.  \end{alignat*} 

Substitute this and then \eqref{eq:djj-dqq} into \eqref{eq:omMMjs} to get
\begin{alignat*}{5} 1-\Mach^2 \creq \frac{\pp_0}{\pp} \cot\bet \cot\bet_0  .\myeqlab{eq:omMM-tan} \end{alignat*} 
Right away we note that the right-hand side is positive, so:
\begin{alignat*}{5}
\text{Critical-type shocks of ideal polars are transonic} \myeqlab{eq:crit-transonic}
\end{alignat*}
(under the assumptions of section \ref{section:assumptions}). In particular they all lie on the one subsonic segment. 

On the right-hand side of \myeqref{eq:omMM-tan}, as $\bet_0$ increases $1/\pp$ is \strictly decreasing \myeqref{eq:genmon}; $\bet$ is \strictly increasing on the subsonic segments \myeqref{eq:bet-incr-sub}, so both $\cot$ are \strictly decreasing, 
and so is the entire right-hand side.
On the other hand the left-hand side is clearly \strictly increasing on the sonic-subsonic part where $\Mach$ is \strictly decreasing \myeqref{eq:Mach-decr-sub}.
Hence the equation cannot hold in more than one $\bet_0$:
\begin{alignat*}{5}
\text{Critical-type shocks of ideal polars are unique} \myeqlab{eq:crit-unique}
\end{alignat*}
(on each half of the polar, under the assumptions of section \ref{section:assumptions}).

\section{Non-ideal polars}
\label{section:nonideal}

We immediately note that uniqueness of critical-type shocks cannot be generalized to non-ideal polars 
if only the fundamental assumptions in section \ref{section:assumptions} are imposed. 
The right-hand side of \eqref{eq:omMM-tan} is a function of coordinates $\pp,\spm$ and up to first derivatives of $\hpm(\pp,\spm)$, via $(\jn)^2=[\pp]/[-\idens]$, $\idens=\dat\hpm\pp\spm$ etc. But the left-hand side 
\begin{alignat*}{5} 1-\Mach^2 = 1-\frac{(\dens\ua)^2}{(\dens\csnd)^2} = 1+\hpm_{\pp\pp}\ja^2 \myeqlab{eq:omMM-h} \end{alignat*}
involves a second derivative $\hpm_{\pp\pp}$. So monotonicity of the right-hand side involves at most second derivatives, while the left-hand side requires some third ones, 
namely those in
\begin{alignat*}{5}
\dehug(\hpm_{\pp\pp}) = \hpm_{\pp\pp\pp} \dehug\pp + \hpm_{\pp\pp\spm} \dehug\spm . \end{alignat*}
$\hpm_{\pp\pp\pp}$ is constrained by convex equation of state \eqref{eq:hppp-gennon},
but $\hpm_{\pp\pp\spm}$ is free. 
It is possible to modify any $m$th partial derivative of a smooth function in an arbitrarily small region by an arbitrarily large amount in either direction, while making arbitrarily small changes to the other $m$th and lower derivatives. 
So we can add strong oscillation to the left-hand side of \eqref{eq:omMM-h} without changing the right-hand side much,
so that they equal each other any number of times.\footnote{In the ideal case $\pp\hpm_\pp=\pp\idens=\stemp=\hpm_\spm$ prevents changing 3rd derivatives individually.}

On the other hand the non-ideal polars observed in many concrete applications are usually still convex in the $\ux,\uy$ plane, and otherwise well-behaved.

Some prior works (\cite{teshukov-polar}, \cite[Appendix B]{henderson-menikoff} 
who also attribute some results to \cite{fowles-jfm1981}) have given various equivalent or sufficient conditions 
for subsonicness or uniqueness of critical-type shocks; 
these conditions are implicit in the sense that additional verification of non-obvious properties is needed, 
for example verifying monotonicity of $\hpm+\half\csnd^2$ 
or comparison of $\jmp\pp/\pp$ to $\csnd^2/\pp\idens$ and $\dat\pp\spm\idens$ along the Hugoniot curve. 
These conditions cannot be trivial; 
e.g.\ we have already seen $\csnd$ monotonicity need not hold even for ideal gas, 
while various inequalities for the Gr\"uneisen coefficient 
\begin{alignat*}{5} \Grun = \dat{\ln\stemp}{\ln\dens}\spm = - \frac{ \hpm_\pp\hpm_{\pp\spm} }{ \hpm_\spm \hpm_{\pp\pp} }  \end{alignat*}
imposed in these articles only constrain up to second derivatives of $\hpm$, not $\hpm_{\pp\pp\spm}$ (or $\hpm_{\pp\spm\spm},\hpm_{\spm\spm\spm}$). 

The only route for progress is to focus on particular classes of realistic gases, 
trying to find reasonable conditions that rule out pathological behaviour.

\section*{Acknowledgement}

This research was partially supported by Taiwan MOST Grant No.\ 108-2115-M-001-002-MY2.

\bibliographystyle{amsalpha}
\input{idealpolar.bbl}

\end{document}

%% file: bluntbodyscaled.pdf_t
\begin{picture}(0,0)%
\includegraphics{bluntbodyscaled.pdf}%
\end{picture}%
\setlength{\unitlength}{3947sp}%
\begingroup\makeatletter\ifx\SetFigFont\undefined%
\gdef\SetFigFont#1#2#3#4#5{%
  \reset@font\fontsize{#1}{#2pt}%
  \fontfamily{#3}\fontseries{#4}\fontshape{#5}%
  \selectfont}%
\fi\endgroup%
\begin{picture}(2036,1599)(7786,-5023)
\put(9315,-4426){\makebox(0,0)[lb]{\smash{{\SetFigFont{6}{7.2}{\rmdefault}{\mddefault}{\updefault}{\color[rgb]{0,.69,0}supersonic}%
}}}}
\put(7801,-3661){\makebox(0,0)[lb]{\smash{{\SetFigFont{6}{7.2}{\rmdefault}{\mddefault}{\updefault}{\color[rgb]{0,0,0}$\uu_0$}%
}}}}
\put(8583,-4251){\makebox(0,0)[b]{\smash{{\SetFigFont{6}{7.2}{\rmdefault}{\mddefault}{\updefault}{\color[rgb]{0,0,0}solid}%
}}}}
\put(8499,-4643){\makebox(0,0)[rb]{\smash{{\SetFigFont{6}{7.2}{\rmdefault}{\mddefault}{\updefault}{\color[rgb]{0,0,0}shock}%
}}}}
\put(8395,-4555){\makebox(0,0)[rb]{\smash{{\SetFigFont{6}{7.2}{\rmdefault}{\mddefault}{\updefault}{\color[rgb]{0,0,0}bow}%
}}}}
\put(8701,-4486){\makebox(0,0)[lb]{\smash{{\SetFigFont{6}{7.2}{\rmdefault}{\mddefault}{\updefault}{\color[rgb]{0,.69,0}subsonic}%
}}}}
\end{picture}%

%% file: wedge.pdf_t
\begin{picture}(0,0)%
\includegraphics{wedge.pdf}%
\end{picture}%
\setlength{\unitlength}{3947sp}%
\begingroup\makeatletter\ifx\SetFigFont\undefined%
\gdef\SetFigFont#1#2#3#4#5{%
  \reset@font\fontsize{#1}{#2pt}%
  \fontfamily{#3}\fontseries{#4}\fontshape{#5}%
  \selectfont}%
\fi\endgroup%
\begin{picture}(1824,1824)(-11,-973)
\put(1188,-27){\makebox(0,0)[lb]{\smash{{\SetFigFont{6}{7.2}{\rmdefault}{\mddefault}{\updefault}{\color[rgb]{0,0,0}$\turn$}%
}}}}
\put(1133,-175){\makebox(0,0)[lb]{\smash{{\SetFigFont{6}{7.2}{\rmdefault}{\mddefault}{\updefault}{\color[rgb]{0,0,0}$\turn$}%
}}}}
\put(1521,-236){\makebox(0,0)[lb]{\smash{{\SetFigFont{6}{7.2}{\rmdefault}{\mddefault}{\updefault}{\color[rgb]{0,0,0}solid}%
}}}}
\put( 62,432){\makebox(0,0)[lb]{\smash{{\SetFigFont{6}{7.2}{\rmdefault}{\mddefault}{\updefault}{\color[rgb]{0,0,0}$\uu_0$}%
}}}}
\put(1209,300){\rotatebox{29.5}{\makebox(0,0)[b]{\smash{{\SetFigFont{6}{7.2}{\rmdefault}{\mddefault}{\updefault}{\color[rgb]{.63,.25,0}weak shock}%
}}}}}
\put(1363,187){\rotatebox{15.0}{\makebox(0,0)[lb]{\smash{{\SetFigFont{6}{7.2}{\rmdefault}{\mddefault}{\updefault}{\color[rgb]{.63,.25,0}$\uu$}%
}}}}}
\put(1302,-455){\rotatebox{345.0}{\makebox(0,0)[lb]{\smash{{\SetFigFont{6}{7.2}{\rmdefault}{\mddefault}{\updefault}{\color[rgb]{1,0,0}$\uu$}%
}}}}}
\put(785,-513){\rotatebox{289.0}{\makebox(0,0)[b]{\smash{{\SetFigFont{6}{7.2}{\rmdefault}{\mddefault}{\updefault}{\color[rgb]{1,0,0}strong shock}%
}}}}}
\end{picture}%

%% file: shockbet.pdf_t
\begin{picture}(0,0)%
\includegraphics{shockbet.pdf}%
\end{picture}%
\setlength{\unitlength}{3947sp}%
\begingroup\makeatletter\ifx\SetFigFont\undefined%
\gdef\SetFigFont#1#2#3#4#5{%
  \reset@font\fontsize{#1}{#2pt}%
  \fontfamily{#3}\fontseries{#4}\fontshape{#5}%
  \selectfont}%
\fi\endgroup%
\begin{picture}(2864,2202)(-382,-1907)
\put(751,-1636){\makebox(0,0)[lb]{\smash{{\SetFigFont{8}{9.6}{\rmdefault}{\mddefault}{\updefault}{\color[rgb]{0,0,0}shock}%
}}}}
\put(-299,-1861){\makebox(0,0)[lb]{\smash{{\SetFigFont{8}{9.6}{\rmdefault}{\mddefault}{\updefault}{\color[rgb]{0,0,0}upstream}%
}}}}
\put(-299,-1711){\makebox(0,0)[lb]{\smash{{\SetFigFont{8}{9.6}{\rmdefault}{\mddefault}{\updefault}{\color[rgb]{0,0,0}$\dens_0=1$}%
}}}}
\put(1377,-1376){\makebox(0,0)[lb]{\smash{{\SetFigFont{8}{9.6}{\rmdefault}{\mddefault}{\updefault}{\color[rgb]{0,0,1}normal}%
}}}}
\put(1341,-985){\makebox(0,0)[rb]{\smash{{\SetFigFont{8}{9.6}{\rmdefault}{\mddefault}{\updefault}{\color[rgb]{0,0,0}$\turn$}%
}}}}
\put(1426,-1711){\makebox(0,0)[lb]{\smash{{\SetFigFont{8}{9.6}{\rmdefault}{\mddefault}{\updefault}{\color[rgb]{0,0,0}$\dens>1$}%
}}}}
\put(1426,-1861){\makebox(0,0)[lb]{\smash{{\SetFigFont{8}{9.6}{\rmdefault}{\mddefault}{\updefault}{\color[rgb]{0,0,0}downstream}%
}}}}
\put(1009,-1097){\makebox(0,0)[b]{\smash{{\SetFigFont{8}{9.6}{\rmdefault}{\mddefault}{\updefault}{\color[rgb]{0,0,0}$\bet$}%
}}}}
\put(  1,-1186){\rotatebox{32.0}{\makebox(0,0)[lb]{\smash{{\SetFigFont{8}{9.6}{\rmdefault}{\mddefault}{\updefault}{\color[rgb]{0,0,0}$\uu_0=\jj_0$}%
}}}}}
\put(1959,-141){\rotatebox{45.0}{\makebox(0,0)[rb]{\smash{{\SetFigFont{8}{9.6}{\rmdefault}{\mddefault}{\updefault}{\color[rgb]{1,0,0}$\jj$}%
}}}}}
\put(1488,208){\makebox(0,0)[b]{\smash{{\SetFigFont{6}{7.2}{\rmdefault}{\mddefault}{\updefault}{\color[rgb]{1,0,0}$\jj\dotp\nn=\jj_0\dotp\nn$}%
}}}}
\put(2176,-511){\rotatebox{90.0}{\makebox(0,0)[lb]{\smash{{\SetFigFont{6}{7.2}{\rmdefault}{\mddefault}{\updefault}{\color[rgb]{1,0,0}tangent}%
}}}}}
\put(2441,-912){\rotatebox{90.0}{\makebox(0,0)[b]{\smash{{\SetFigFont{6}{7.2}{\rmdefault}{\mddefault}{\updefault}{\color[rgb]{0,0,1}$\uu\dotp\ts=\uu_0\dotp\ts$}%
}}}}}
\put(2092,-640){\rotatebox{32.0}{\makebox(0,0)[rb]{\smash{{\SetFigFont{6}{7.2}{\rmdefault}{\mddefault}{\updefault}{\color[rgb]{0,0,0}$\uu_0{=}\jj_0$}%
}}}}}
\put(1523,-601){\rotatebox{45.0}{\makebox(0,0)[rb]{\smash{{\SetFigFont{6}{7.2}{\rmdefault}{\mddefault}{\updefault}{\color[rgb]{0,0,1}$\uu$}%
}}}}}
\put(1111,-854){\makebox(0,0)[b]{\smash{{\SetFigFont{8}{9.6}{\rmdefault}{\mddefault}{\updefault}{\color[rgb]{0,0,0}$\bet_0$}%
}}}}
\end{picture}%

%% file: deltajq.pdf_t
\begin{picture}(0,0)%
\includegraphics{deltajq.pdf}%
\end{picture}%
\setlength{\unitlength}{3947sp}%
\begingroup\makeatletter\ifx\SetFigFont\undefined%
\gdef\SetFigFont#1#2#3#4#5{%
  \reset@font\fontsize{#1}{#2pt}%
  \fontfamily{#3}\fontseries{#4}\fontshape{#5}%
  \selectfont}%
\fi\endgroup%
\begin{picture}(3870,2162)(-11,-1112)
\put(1241,-343){\makebox(0,0)[b]{\smash{{\SetFigFont{6}{7.2}{\rmdefault}{\mddefault}{\updefault}{\color[rgb]{0,0,0}$\uu$}%
}}}}
\put(1173,-507){\rotatebox{24.0}{\makebox(0,0)[b]{\smash{{\SetFigFont{6}{7.2}{\rmdefault}{\mddefault}{\updefault}{\color[rgb]{0,0,1}$\delta\uu$}%
}}}}}
\put(284,-936){\makebox(0,0)[b]{\smash{{\SetFigFont{6}{7.2}{\rmdefault}{\mddefault}{\updefault}{\color[rgb]{0,0,0}$\turn$}%
}}}}
\put(3035,504){\rotatebox{24.0}{\makebox(0,0)[b]{\smash{{\SetFigFont{6}{7.2}{\rmdefault}{\mddefault}{\updefault}{\color[rgb]{0,0,1}$\delta\jj$}%
}}}}}
\put(3198,450){\makebox(0,0)[b]{\smash{{\SetFigFont{6}{7.2}{\rmdefault}{\mddefault}{\updefault}{\color[rgb]{0,0,0}$\jj$}%
}}}}
\put(2407,-1070){\makebox(0,0)[b]{\smash{{\SetFigFont{6}{7.2}{\rmdefault}{\mddefault}{\updefault}{\color[rgb]{0,0,0}$\uu_0=\jj_0$}%
}}}}
\put(2711,-864){\makebox(0,0)[b]{\smash{{\SetFigFont{6}{7.2}{\rmdefault}{\mddefault}{\updefault}{\color[rgb]{0,0,0}$\bet_0$}%
}}}}
\put(2992, 28){\rotatebox{62.0}{\makebox(0,0)[b]{\smash{{\SetFigFont{6}{7.2}{\rmdefault}{\mddefault}{\updefault}{\color[rgb]{0,0,0}shock}%
}}}}}
\put(697,-192){\rotatebox{332.0}{\makebox(0,0)[b]{\smash{{\SetFigFont{6}{7.2}{\rmdefault}{\mddefault}{\updefault}{\color[rgb]{0,0,0}normal}%
}}}}}
\put(3350,689){\makebox(0,0)[b]{\smash{{\SetFigFont{6}{7.2}{\rmdefault}{\mddefault}{\updefault}{\color[rgb]{0,0,0}$\bet$}%
}}}}
\end{picture}%

%% file: jqlin.pdf_t
\begin{picture}(0,0)%
\includegraphics{jqlin.pdf}%
\end{picture}%
\setlength{\unitlength}{3947sp}%
\begingroup\makeatletter\ifx\SetFigFont\undefined%
\gdef\SetFigFont#1#2#3#4#5{%
  \reset@font\fontsize{#1}{#2pt}%
  \fontfamily{#3}\fontseries{#4}\fontshape{#5}%
  \selectfont}%
\fi\endgroup%
\begin{picture}(2463,1435)(553,-4166)
\put(1214,-3850){\makebox(0,0)[lb]{\smash{{\SetFigFont{8}{9.6}{\rmdefault}{\mddefault}{\updefault}{\color[rgb]{0,0,0}subsonic}%
}}}}
\put(676,-2761){\rotatebox{90.0}{\makebox(0,0)[rb]{\smash{{\SetFigFont{8}{9.6}{\rmdefault}{\mddefault}{\updefault}{\color[rgb]{0,0,0}mass flux $\ja$}%
}}}}}
\put(1801,-2836){\makebox(0,0)[rb]{\smash{{\SetFigFont{8}{9.6}{\rmdefault}{\mddefault}{\updefault}{\color[rgb]{0,0,0}critical}%
}}}}
\put(2701,-2836){\makebox(0,0)[rb]{\smash{{\SetFigFont{8}{9.6}{\rmdefault}{\mddefault}{\updefault}{\color[rgb]{0,0,0}limit}%
}}}}
\put(3001,-4111){\makebox(0,0)[rb]{\smash{{\SetFigFont{8}{9.6}{\rmdefault}{\mddefault}{\updefault}{\color[rgb]{0,0,0}speed $\ua$}%
}}}}
\put(1848,-3853){\makebox(0,0)[lb]{\smash{{\SetFigFont{8}{9.6}{\rmdefault}{\mddefault}{\updefault}{\color[rgb]{0,0,0}supersonic}%
}}}}
\end{picture}%

%% file: idealpolar.bbl
\providecommand{\bysame}{\leavevmode\hbox to3em{\hrulefill}\thinspace}
\providecommand{\MR}{\relax\ifhmode\unskip\space\fi MR }
\providecommand{\MRhref}[2]{%
  \href{http://www.ams.org/mathscinet-getitem?mr=#1}{#2}
}
\providecommand{\href}[2]{#2}